\documentclass[12pt]{article}

\usepackage[english]{babel}

\usepackage[letterpaper,top=2cm,bottom=2cm,left=3cm,right=3cm,marginparwidth=1.75cm]{geometry}

\usepackage{amsmath}
\usepackage{graphicx}
\usepackage[colorlinks=true, allcolors=blue]{hyperref}

\title{There is no electroweak horizon problem}
\author{James M.\ Cline}
\date{% 
{\it McGill University, Dept.\ of Physics, Montr\'eal, Qu\'ebec, Canada}\\
}

\begin{document}
\maketitle

\begin{abstract}
Contrary to a recent assertion, there is no electroweak horizon problem.
\end{abstract}

It was recently claimed \cite{stupid} that the electroweak phase transition  (EWPT) in the early universe leads to a horizon problem, since (according to the claim)
there is no reason for the vacuum expectation value (VEV)  $v$ of the Higgs field
 to take the same value in different causally disconnected regions of space.
This is based on the mistaken assertion, not explained nor justified, that 
$v$ ``could have been anything'' in different regions, and ``we have no reason to believe that the VEV is specified uniquely.''

In fact, we have a very good reason, namely that the Higgs field has a potential $V(H) = \lambda(|H|^2-v^2)^2$, where $v$ is a fixed constant of nature, not a spatially varying modulus.  Therefore the VEV of the Higgs field is dynamically determined to be $v=246$\,GeV today, as well as shortly after the EWPT.  The situation where the Higgs VEV $\langle|H|\rangle$ would be undetermined is that in which the potential is flat and the Higgs field is
massless, which is not what has been observed \cite{ATLAS:2012yve,CMS:2012qbp}.  So far, all indications from collider physics are that the Higgs potential is consistent with the standard model (SM) prediction.

Suppose, for the sake of argument, that it were somehow possible to arrange for $\langle |H|\rangle$ to deviate from its SM value in some region of space.  This would be equivalent to a condensate of zero-momentum Higgs bosons with energy density $V(\langle |H|\rangle)$.
These particles would decay into other SM particles, driving $\langle |H|\rangle\to v$ on the time scale $\hbar /(4\,{\rm MeV}) \sim 10^{-22}\,$s.
Therefore, if it were possible to engineer an electroweak horizon problem, it would rather quickly resolve itself.  Within the particle physics community, all of these statements are perfectly obvious and well understood, with no margin for controversy.

\bibliographystyle{unsrt}
\bibliography{sample}

\end{document}